# Large elasto-optic effect and reversible electrochromism in multiferroic $BiFeO_3$

D. Sando[1,*,†], Yurong Yang[2,*], E. Bousquet[3], C. Carrétéro[1], V. Garcia[1], S. Fusil[1], D. Dolfi[4], A. Barthélémy[1], Ph. Ghosez[3], L. Bellaiche[2] & M. Bibes[1]

The control of optical fields is usually achieved through the electro-optic or acousto-optic effect in single-crystal ferroelectric or polar compounds such as $LiNbO_3$ or quartz. In recent years, tremendous progress has been made in ferroelectric oxide thin film technology—a field which is now a strong driving force in areas such as electronics, spintronics and photovoltaics. Here, we apply epitaxial strain engineering to tune the optical response of $BiFeO_3$ thin films, and find a very large variation of the optical index with strain, corresponding to an effective elasto-optic coefficient larger than that of quartz. We observe a concomitant strain-driven variation in light absorption—reminiscent of piezochromism—which we show can be manipulated by an electric field. This constitutes an electrochromic effect that is reversible, remanent and not driven by defects. These findings broaden the potential of multiferroics towards photonics and thin film acousto-optic devices, and suggest exciting device opportunities arising from the coupling of ferroic, piezoelectric and optical responses.

[1] Unité Mixte de Physique, CNRS, Thales, Univ. Paris-Sud, Université Paris-Saclay, 91767 Palaiseau, France. [2] Department of Physics and Institute for Nanoscience and Engineering, University of Arkansas, Fayetteville, Arkansas 72701, USA. [3] Theoretical Materials Physics, Université de Liège, B-5, B-4000 Sart-Tilman, Belgium. [4] Thales Research and Technology France, 1 Avenue Augustin Fresnel, 91767 Palaiseau, France. * These authors contributed equally to this work. † Present address: School of Materials Science and Engineering, University of New South Wales, Sydney 2052, Australia. Correspondence and requests for materials should be addressed to D.S. (email: daniel.sando@unsw.edu.au) or to M.B. (email: manuel.bibes@thalesgroup.com).

**B**ismuth ferrite (BiFeO$_3$—BFO) is multiferroic at room temperature with strong ferroelectric polarisation[1] and G-type antiferromagnetic ordering with a cycloidal modulation of the Fe spins[2]. Most research on this material has been driven by the prospect of electrically controlled spintronic devices[3]. More recently, however, BFO has revealed further remarkable multifunctional properties. Notable discoveries include conductive domain walls[4], a strain-driven morphotropic phase boundary[5] and a specific magnonic response that can be tuned by epitaxial strain[6] or electric field[7]. Moreover, with a bandgap ($\sim 2.7$ eV) in the visible[8], large birefringence[9] (0.25–0.3), a strong photovoltaic effect[10] and sizeable linear electro-optic coefficients[11], BFO is garnering interest in photonics and plasmonics[12].

Most of these physical properties are intimately linked to structural parameters, and may thus be tuned in thin films by epitaxial strain. Strain engineering[13] is a powerful tool through which, for instance, ferroelectricity is strongly enhanced in BaTiO$_3$ (ref. 14), or induced in otherwise non-ferroelectric materials such as SrTiO$_3$ (ref. 15). In BFO, two structural instabilities are sensitive to epitaxial strain: the polar distortion—responsible for the ferroelectricity—and antiferrodistortive (FeO$_6$ octahedra) rotations. In strained BFO films, the competition between both instabilities and their coupling to ferroic order parameters yields rich phase diagrams, revealing new structural, ferroelectric and magnetic phases[6,16], as well as large variations in the ferroelectric Curie temperature[17] and the spin direction[6].

Here, we present a combined experimental and theoretical study demonstrating that strain induces a very large change in the refractive index of BFO, which corresponds to an effective elasto-optic coefficient larger than in any ferroelectric, and larger than that of quartz[18]. This effect is accompanied by a shift of the optical bandgap, reminiscent of pressure-induced changes in light-absorption[19], a phenomenon known as piezochromism in other materials systems[20]. The trends in the optical properties as a function of strain are well reproduced by our first-principles calculations, and we are able to clarify precisely why the optical bandgap of tetragonal-like BFO is larger than that of the rhombohedral-like phase. Finally, we show how an electric field can be used to toggle between two strain states with different light absorption, corresponding to an electrochromic effect that is intrinsic, reversible and non-volatile.

## Results

**Sample preparation and structural characterization**. Fully strained BFO thin films were grown using pulsed laser deposition on (001)-oriented substrates (in pseudocubic notation, which we use throughout this paper) spanning a broad range of lattice mismatch (from $-7.0\%$ to $+1.0\%$; Methods section). At low strain—compressive or tensile—the films crystallize in the so-called R-like phase of BFO, derived from the bulk rhombohedral (R3c) phase. At high compressive strain ($\leq 4\%$), the films grow in the T-like phase[16] with a large tetragonality ratio $c/a \approx 1.26$ (cf. Fig. 1b). Reciprocal space maps around the $(113)_{pc}$ or $(223)_{pc}$ reflections (Supplementary Fig. 1) reveal that all our BFO films possess a monoclinic structure (M$_A$ or M$_B$ for R-like, M$_C$ for T-like[12,21]), and further scans (not shown) indicate the presence of two structural domain variants (see the sketch in Fig. 1a). The in-plane and out-of-plane pseudocubic lattice parameters are presented in Fig. 1b. The in-plane parameter shows a monotonic decrease with compressive strain, while the out-of-plane parameter concomitantly increases, albeit with a sharp jump at $\sim -3.5\%$ corresponding to the structural transition between the R-like and T-like phases.

**First-principles calculations**. To explore the effect of strain on the optical properties of BFO thin films, we performed first-principles calculations, using the Heyd–Scuseria–Ernzerhof (HSE) hybrid functional (see Methods section for details). In the following, we denote the electronic bandgap as the energy difference between the valence band maximum (VBM) and the conduction band minimum (CBM), while the optical bandgap corresponds to the extrapolation of the linear region of the Tauc plot (cf. Fig. 2c); theoretically, the optical bandgap is computed from the complex dielectric function.

Figure 1c,d show the computed electronic density of states for the R-like phase of BFO (at 0, $+2$ and $-3\%$ strain) and for the T-like phase ($-5$ and $-7\%$). The insets of Fig. 1c,d show that the electronic bandgap is lower for the T-like phase than for the R-like phase, particularly for $-7\%$ strain, consistent with previous studies[22,23]. In the R-like phase, both compressive and tensile strains yield an increase of the electronic bandgap, similar to the situation in SrTiO$_3$ (ref. 24).

The partial density of states of Fig. 1e-l shows that the VBM mostly consists of O 2p orbitals for any considered misfit strain, and the CBM mainly comprises Fe $d_{xy}$, $d_{xz}$, $d_{yz}$ orbitals for both the R-like and T-like phases. In R-like BFO, strain-induced changes in the FeO$_6$ octahedra rotations and the polar modes conspire[24] to slightly lift the degeneracy of the Fe 3d orbitals, but the nature of the electronic states at the VBM and CBM is globally preserved. In contrast, the pyramidal coordination of the FeO$_5$ unit in highly elongated T-like BFO yields a large splitting of the 3d states with the $d_{xy}$ orbital sitting 300 meV lower in energy (Fig. 1h). Figure 1l shows that this $d_{xy}$ state is weakly hybridized with O states and the states near the CBM in the T-like phase have very little O 2p character. This suggests that optical transitions from the VBM to those states should be very weak, and that the main optical transitions in the T-like phase should occur from the VBM to $d_{xz}$ and $d_{yz}$ states that lie in energy $\sim 300$ meV above the CBM. In other words, the optical bandgap should be higher in T-like BFO than in R-like BFO despite the opposite trend in the electronic bandgap. This presumption is confirmed by the energy dependence of the extinction coefficient derived from our calculations (Fig. 2b): the absorption edge appears at least 200 meV higher in the T-like phase than in the R-like phase.

**Optical characterization**. Figure 2a presents the experimental energy dependence of the extinction coefficient, extracted from spectroscopic ellipsometry measurements (Supplementary Fig. 2). These data confirm the theoretical prediction of a larger optical bandgap for the T-like phase, and are consistent with previous studies. The agreement between the experimental and calculated extinction coefficient curves (Fig. 2a,b) is very good, particularly for the onset of absorption. The corresponding experimental Tauc plots (cf. Methods section) for these three samples are shown in Fig. 2c, indicating that the optical bandgap for T-like BFO is 3.02 eV, while for R-like BFO, a compressive strain of 2.6% induces an increase in the bandgap from 2.76 to 2.80 eV. Figure 2d summarizes the strain dependence of the experimental and calculated optical bandgap. In the R-like phase, both compressive and tensile strains induce an increase of the optical bandgap, and the T-like polymorph exhibits an optical bandgap $\sim 0.25$ eV larger than the R-like phase, consistent with previous reports[25].

Our observation of a strain-induced change in optical band-gap and thus optical absorption is reminiscent of an effect called piezochromism[20], which corresponds to changes in light absorption driven by hydrostatic pressure. Piezochromic effects have been identified in several organic compounds, but for inorganics mainly in the CuMoO$_4$ family[26,27]. In this compound, the application of hydrostatic pressure triggers a first order transition between two polytypes having different optical absorption spectra due to changes in the oxygen cage surrounding the Cu ions[26]. Interestingly, in both CuMoO$_4$ and BiFeO$_3$ the absorption is stronger when the transition metal cation is in an octahedral oxygen environment, which suggests a possible trend, and strategies for engineering piezochromic effects in other perovskites.

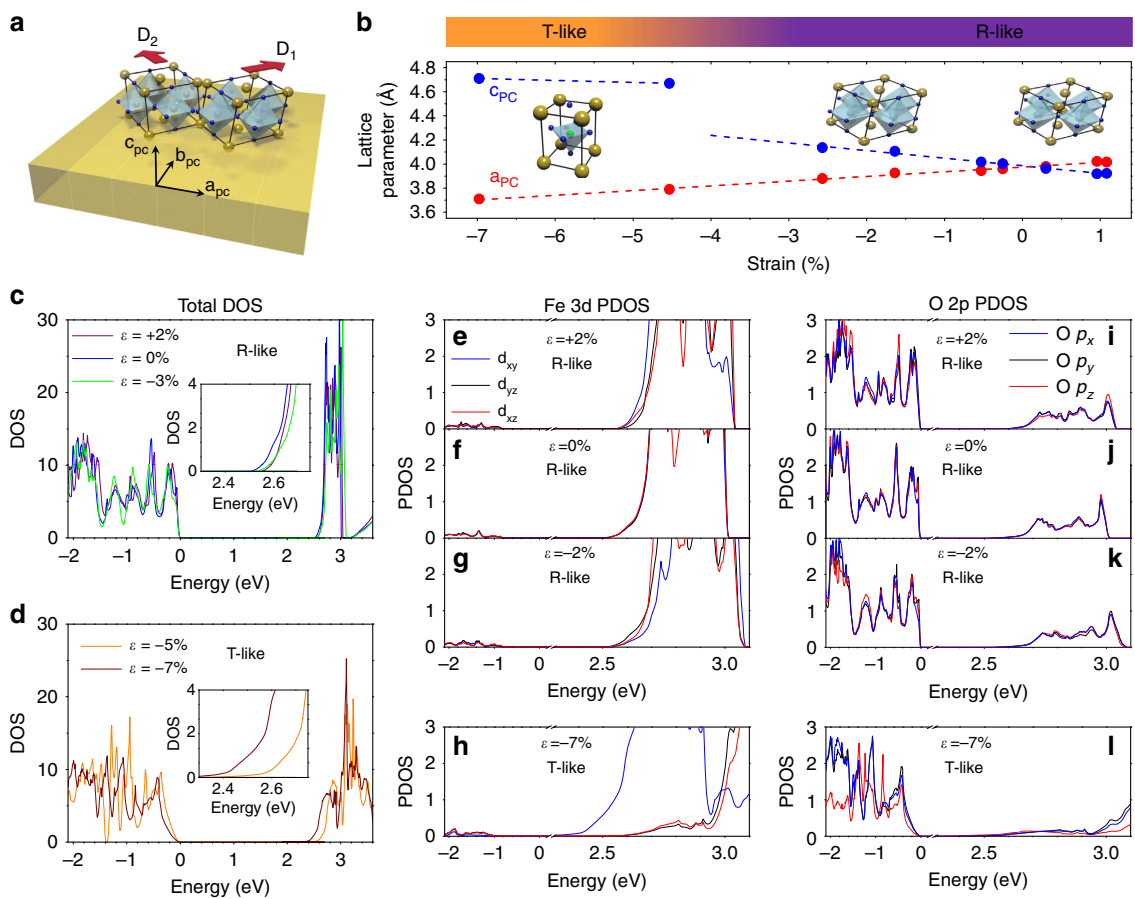

**Figure 1 | Electronic structure of strained BiFeO$_3$ thin films.** (**a**) Sketch of the two structural variants present in our monoclinic (M$_A$ or M$_B$) R-like BiFeO$_3$ films. The red arrows indicate the direction of the monoclinic distortion for the two variants D$_1$ and D$_2$. (**b**) In-plane and out-of-plane lattice parameters of our strained BFO films. Total density of state (DOS) for R-like (**c**) and T-like phases (**d**). The insets show the DOS near the CBM. Partial density of states (PDOS) of iron 3d (**e–h**) and oxygen (**i–l**) states for R-like and T-like BFO. Note the break between 0.2 and 2.3 eV in the horizontal axes in (**e–l**). For all panels, only the spin-up channel states are shown; the spin-down channel states are the same as the spin up due to the antiferromagnetic order.

In bulk BFO, the bandgap is known to decrease with pressure[19], particularly below 3.4 GPa and at the structural phase transition near 9.5 GPa. This corresponds to a piezochromic effect of amplitude 0.058 eV GPa$^{-1}$ at low pressure, and of 0.027 eV GPa$^{-1}$ on average between ambient pressure and 18 GPa (ref. 19). In our films, from the strain values and Young's modulus[28], we estimate the amplitude of the piezochromic effect in BiFeO$_3$ at ∼0.12 eV GPa$^{-1}$. Importantly, working with thin films may be advantageous for several applications[29]. For instance, the thin film geometry allows the application of large electric fields to toggle between two optical polytypes, thereby producing potentially high-speed electrochromic effects.

We have explored this possibility in BFO thin films with coexisting R-like and T-like regions[5]. We applied an electric field to transform the mixed R − T BFO into nominally pure T-like BFO over 10 × 10 μm$^2$ regions (cf. Fig. 3a) and probed the local optical transmission in and out of this area using a conventional optical microscope. Figure 3b shows a transmission image with a dielectric filter (bandwidth 10 nm) centred at 420 nm inserted between the white light source and the sample. Clearly, a 10 × 10 μm$^2$ square with a higher intensity than the background is visible in the image. Remarkably, this effect is reversible: applying a voltage with the opposite polarity restores a mixed R + T state (cf. Fig. 3c), which restores a stronger optical absorption, see Fig. 3d. This contrast is stable for several weeks.

We have acquired similar optical images using various dielectric filters, recorded the transmitted intensity in and out of the 10 × 10 μm$^2$ square and calculated the contrast difference as a function of wavelength. The contrast is maximal between 420 and 450 nm, see Fig. 3e. This dependence agrees very well with the expected contrast difference, calculated from the extinction coefficients of pure R-like BiFeO$_3$ and pure T-like BiFeO$_3$ films of Fig. 2a. This confirms that the contrast in the optical images is indeed due to the intrinsic modulation of the optical bandgap induced by the electrical poling, rather than by defect-mediated processes as in Ca-doped BFO (ref. 30) or WO$_3$ (ref. 20).

Finally, we focus on the influence of strain on the real part of the complex refractive index n. In Fig. 4a, we highlight representative results of the variation of n with wavelength for R-like BFO that is weakly strained (on SmScO$_3$, SSO), strongly compressively strained (on (La,Sr)(Al,Ta)O$_3$ (LSAT)), and T-like BFO (on LaAlO$_3$ (LAO)). Below the optical bandgap (that is, for wavelengths longer than ∼460 nm), the refractive index systematically decreases with increasing strain. This is also visible in

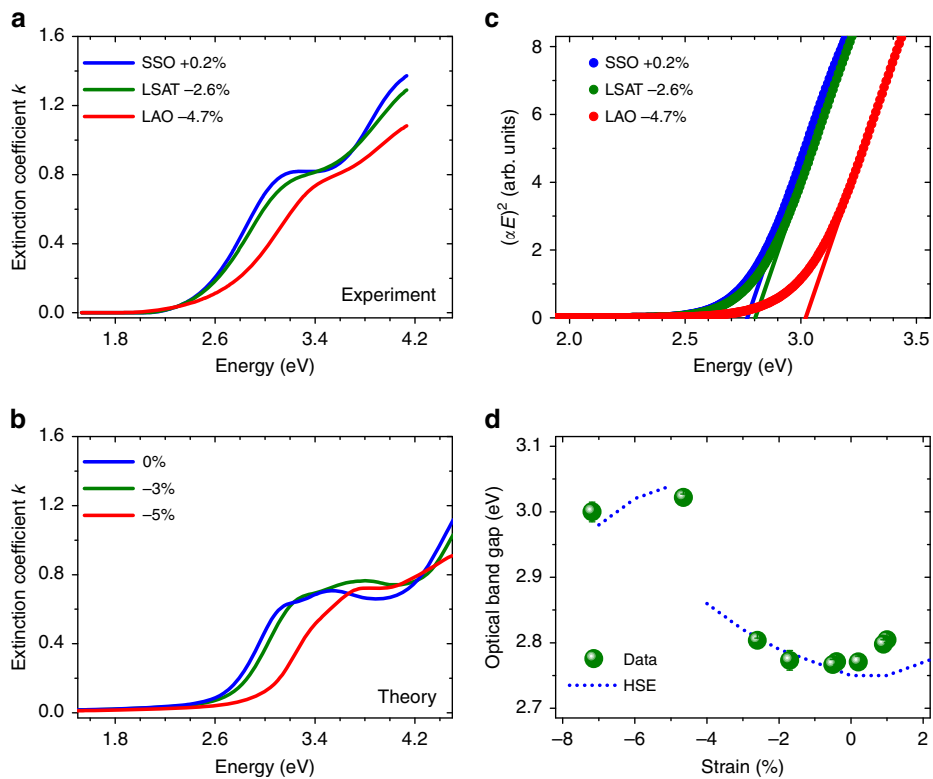

**Figure 2 | Optical absorption properties of strained BiFeO$_3$ films.** (**a**) Measured extinction coefficient for three representative strain levels. (**b**) Calculated extinction coefficient for strain levels comparable to those displayed in **a**. (**c**) Tauc plots generated from measurements for representative samples. (**d**) Summary of optical bandgap versus strain results, comparing theory and experiment. The error bars were determined by generating the dispersion laws using the upper and lower bounds of the Tauc–Lorentz oscillator parameters (from their uncertainties) and finding the resultant maximum variation in the bandgap.

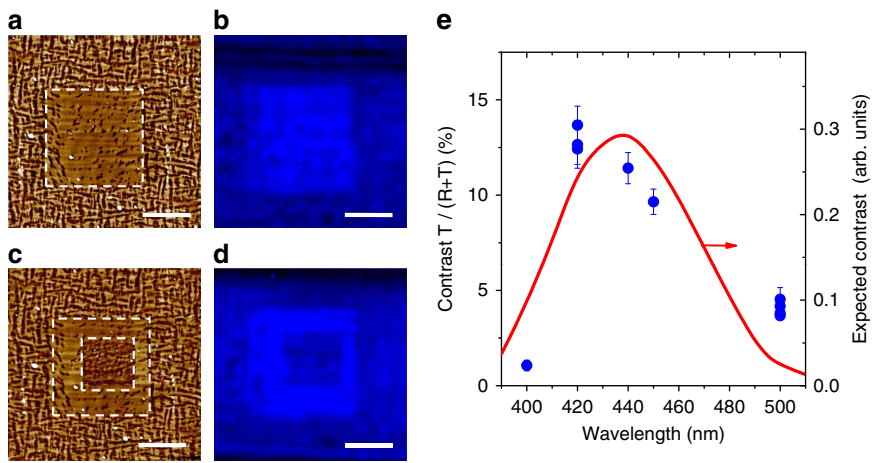

**Figure 3 | Electrochromism in BiFeO$_3$ thin films.** (**a**) Topography image after poling a 10 × 10 μm$^2$ square, locally transforming R + T BFO into T-like BFO. (**b**) Transmission optical image acquired in the same region with a dielectric filter centred at 420 nm (bandwidth 10 nm). (**c**) Topography image of the same area after poling a 5 × 5 μm$^2$ region with an opposite voltage, restoring the R + T structure. (**d**) Transmission optical image with a 420-nm filter. The horizontal dark features are due to twin boundaries in the LaAlO$_3$ substrate. All white scale bars are 5 μm. (**e**) Blue symbols: normalized difference in transmitted light in (T) and out (R + T) of the square in **a** with dielectric filters centred at different wavelengths. Red line: expected contrast calculated from the transmission of pure R-like and T-like films and the transmission function of the dielectric filters. The error bars in **e** are derived from the s.d. of the image pixel values in zones in and out of the T and (R + T) regions.

Fig. 4b, which displays the strain dependence of $n$ at various wavelengths for all samples. The strain-induced change in refractive index measured at 633 nm is reproduced in Fig. 4c and compared with first-principles calculations. The refractive index is higher in the R-like phase and globally decreases with strain, both compressive and tensile.

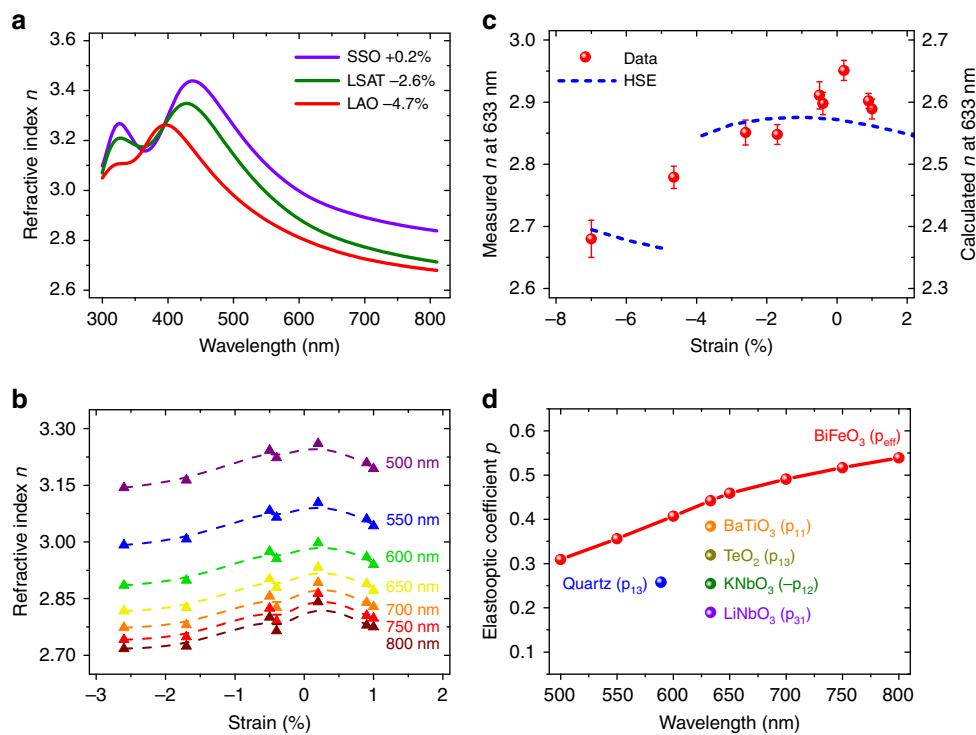

**Figure 4 | Optical refractive index and elasto-optic coefficients in strained BiFeO₃ films.** (**a**) Measured refractive index $n$ as a function of wavelength for various strain levels. (**b**) Measured refractive index as a function of strain for various wavelengths, for the R-like phase only. The lines serve as guides to the eye. (**c**) Measured and calculated refractive index at 633 nm as a function of strain. The error bars were determined by generating the dispersion laws using the upper and lower bounds of the Tauc–Lorentz oscillator parameters (from their uncertainties) and finding the resultant maximum variation in the refractive index. (**d**) Effective elasto-optic coefficient of BFO as a function of wavelength. Representative reported largest elasto-optic coefficients of various other crystalline materials are plotted for comparison.

## Discussion

The experimental results of Fig. 4 indicate that BFO exhibits a strong elasto-optic effect (change in refractive index on physical strain). Taking the slope of the change in $1/n^2$ with strain (Supplementary Fig. 3) for the weakly compressively strained BFO samples at various wavelengths larger than the BFO bandgap (at ~460 nm), we obtain an effective elasto-optic coefficient for BFO, as shown in Fig. 4d (Supplementary Note 1). In this figure we also plot the elasto-optic tensor element with the largest magnitude for various elasto-optic media[18,31]. The results suggest that BFO should be a robust elasto-optic medium and, more specifically, that BFO has an effective elasto-optic coefficient at least twice as large as LiNbO₃.

Combined with its relatively low Young's modulus[28] and sound speed[32], the large elasto-optic coefficient of BFO yields an acousto-optic figure-of-merit[18] $M = \frac{n^6 p^2}{E v_{ac}^3}$ as large as $M = 365 \times 10^{-15}$ s³ kg⁻¹, a value much larger than for any other material for longitudinal acoustic waves (cf. in TeO₂, $M = 23 \times 10^{-15}$ s³ kg⁻¹ and in LiNbO₃ $M = 1.8 \times 10^{-15}$ s³ kg⁻¹, ref. 18). Importantly, as BFO may easily be grown[33] on isostructural perovskites with giant piezoelectric responses (such as PMN-PT or PZN-PT)[34], this huge figure-of-merit opens the way towards thin film acousto-optical components[35] with potential performance orders of magnitude greater than those currently based on single crystals. This would extend the potential of BFO to devices to deflect or modulate optical fields, and towards the emerging field of optomechanics, from back-action and laser cooling, to highly integrated sensors and frequency references[36].

More generally, our work has implications for the design of multifunctional devices exploiting the magnetic, ferroelectric or piezoelectric response of BiFeO₃ in conjunction with these unique optical properties. Importantly, the mechanisms that we identify to modulate bandgap and absorption are not specific to BiFeO₃ and can be transposed to many perovskites. Strain-induced elasto-optic and piezochromic effects even larger than in BiFeO₃, possibly by one order of magnitude or more, could be awaiting discovery in other oxide materials, particularly in Mott insulators[37] in which the bandgap falls between $3d$ states and is lower than in BiFeO₃. Giant, electric field-controllable optical absorption in the visible range could thus be exploited, opening the way towards devices harvesting both mechanical and solar energies.

## Methods

**BiFeO₃ thin film growth and structural characterization.** Single phase R-like or T-like films of BFO were prepared by pulsed laser deposition (using the conditions of ref. 38) on the following single crystal substrates: YAlO₃, LAO, LSAT, SrTiO₃, DyScO₃, TbScO₃, SmScO₃, NdScO₃ and PrScO₃. The scandates and YAlO₃ were (110)-oriented (orthorhombic notation) while cubic SrTiO₃, LSAT and rhombohedral LAO were (001)-oriented. The nominal biaxial strain induced by these substrates ranges from −7.0% (compressive) to +1.0% (tensile). The thickness of the films was 50–70 nm as determined by X-Ray reflectometry, and confirmed by spectroscopic ellipsometry measurements. High-angle X-Ray diffraction $2\theta$–$\theta$ scans, collected with a Panalytical Empyrean diffractometer using CuK$_{\alpha 1}$ radiation, indicated that the films were epitaxial and grew in a single phase. Mixed R + T phase BFO (nominal thickness 100 nm) were grown using a KrF excimer laser at 540 °C and 0.36 mbar on (001)-oriented LAO substrates after the growth of a 10-nm-thick LaNiO₃ bottom electrode at 640 °C and 0.36 mbar.

**Optical characterization.** The films were characterized using spectroscopic ellipsometry with a UVISEL spectral-scanning near infrared spectroscopic phase-modulated ellipsometer from HORIBA Jobin-Yvon. The incidence angle was 70° and the wavelength range was 300–840 nm (0.62–4.13 eV). This range was imposed by the ellipsometer (maximum energy ~4.13 eV), while absorption peaks arising from colour centres in the scandate substrates limited the maximum wavelength to 840 nm. These boundaries do not adversely affect the present analysis since the optical bandgap of BFO is well within the explored spectral range. The raw ellipsometry data were fitted to a multilayer model consisting a semi-infinite substrate, BFO layer and roughness layer implemented by the Bruggeman approximation with a void and BFO mixture. The dispersion law of the BFO layer was described by three Tauc–Lorentz oscillators[8], where the central energies correspond to charge transfer transitions. An example of a typical fit (in this case for BFO on NdScO$_3$) is shown in Supplementary Fig. 2b. The complex dispersion law ($\tilde{n} = n + ik$) of BFO was determined for each sample, and for all fits the mean square error, $\chi^2$, was $< 2$ (as indicated in Supplementary Table 1). To extract the bandgap from the dispersion laws, Tauc plots of $(\alpha E)^2$ versus $E$ were constructed, and the linear region was extrapolated to the $E$ axis (Fig. 2c), yielding the gap value. For each sample, ellipsometric data were collected and the dispersion law and bandgap calculated a minimum of four times, and the results averaged. The error bars displayed in Figs 2d and 4c were determined by generating the dispersion laws using the upper and lower bounds of the Tauc–Lorentz oscillator parameters (from their uncertainties) and finding the resultant maximum variation in the bandgap and refractive index.

**First-principles calculations.** Calculations were performed within density-functional theory, as implemented in the Vienna ab initio simulation package[39,40]. An energy cutoff of 550 eV was used, and the set of projector-augmented wave potentials was employed to describe the electron–ion interaction. We considered the following valence electron configuration: $5d^{10}6s^26p^3$ for Bi, $3p^63d^64s^2$ for Fe and $2s^22p^4$ for O. Supercells containing 20 atoms were used, and G-type antiferromagnetism was adopted. Electronic relaxations converged within $10^{-6}$ eV and ionic relaxation was performed until the residual force was $< 1$ meV Å$^{-1}$. We used the PBEsol + U functional[41] (selecting $U = 4$ eV for the Fe ions) to relax the structures, and used both this PBEsol + U functional and the HSE hybrid functional[42] to calculate physical properties such as electronic structure and the dielectric function. These two methods yielded very similar results (hence we only report results for HSE in Figs 1, 2 and 4), with the exception that the PBEsol + U functional underestimated the electronic bandgap by 0.4 eV, while HSE overestimated this bandgap by 0.8 eV. The imaginary part of the dielectric tensor was obtained via

$$\varepsilon''_{\alpha\beta}(\omega) = \frac{4\pi^2 e^2}{\Omega} \lim_{q \to 0} \frac{1}{q^2} \sum_{c,v,\mathbf{k}} 2\omega_{\mathbf{k}} \delta(\epsilon_{c\mathbf{k}} - \epsilon_{v\mathbf{k}} - \omega) \times \langle u_{c\mathbf{k}+\mathbf{e}_\alpha q} | u_{v\mathbf{k}} \rangle \langle u_{c\mathbf{k}+e_\beta q} | u_{v\mathbf{k}} \rangle^*,$$

(1)

where the indices $c$ and $v$ refer to conduction and valence band states, respectively, $u_{c\mathbf{k}}$ is the cell periodic part of the orbitals at the $k$-point $\mathbf{k}$, and $\mathbf{e}_\alpha$ is a unit vector along the $\alpha$ Cartesian direction[43]. Finally, the real part of the dielectric tensor $\varepsilon'_{\alpha\beta}$ was obtained through the Kramers–Kronig transformation $\varepsilon'_{\alpha\beta}(\omega) = 1 + \frac{2}{\pi} P \int_0^\infty \frac{\varepsilon''_{\alpha\beta}(\omega')\omega'}{\omega'^2 - \omega^2} d\omega'$, where P denotes the principal value. We then obtained the extinction coefficient $k$ and refractive index $n$ by $\tilde{\varepsilon} = \varepsilon' + i\varepsilon'' = (n + ik)^2$. Note that local field effects were neglected in our calculations. The optical bandgap determined from the calculated dielectric function was seen to overestimate the experiment by 0.8 eV; therefore, in all figures in this manuscript, the conduction band has been systematically shifted by 0.8 eV with respect to the VBM to reflect this scissors correction.

For the refractive index we find a systematic quantitative difference of $\sim 0.3$ between experiment and theory, which can be understood by the fact that first-principles calculations consider defect-free samples, neglect local field and temperature effects, and only incorporate the average between the different components of the dielectric function tensor (which is, additionally, a quantity rather difficult to simulate precisely by ab initio methods).

## Acknowledgements

This work was supported by the French Research Agency (ANR) projects 'Méloïc', 'Nomilops' and 'Multidolls,' the European Research Council Advanced Grant 'FEMMES' (Contract No. 267579) and the European Research Council Consolidator Grant 'MINT' (Contract No. 615759). Y.Y. and L.B. thank the financial support of ONR Grant No N00014-12-1-1034 and DARPA Grant No. HR0011-15-2-0038 (Matrix program), and the Arkansas High Performance Computer Center for the use of its supercomputers. Ph.G. acknowledges a Research Professorship from the Francqui Foundation, financial support of the ARC project 'AIMED' and F.R.S.-FNRS PDR project ' HiT4FiT' as well as access to Céci-HPC facilities funded by F.R.S.-FNRS (Grant No 2.5020.1) and the Tier-1 supercomputer of the Fédération Wallonie- Bruxelles funded by the Walloon Region (Grant No 1117545). This work was also supported (E.B.) by F.R.S.-FNRS Belgium, and calculations were partly performed within the PRACE projects TheoMoMuLaM and TheDeNoMo. We thank J.-L. Reverchon for assistance with the ellipsometry measurements.

### Author contributions

M.B. conceived and supervised the study with the help of A.B. and D.D. D.S. and C.C. grew the samples and characterized them with X-ray diffraction. D.S. performed ellipsometry measurements and analysed the data. V.G., S.F. and M.B. characterized the electrochromic response. Y.Y, E.B., Ph.G. and L.B. performed first-principles calculations. D.S. and M.B. wrote the manuscript with input from all authors.

### Additional information

**Supplementary Information** accompanies this paper at http://www.nature.com/naturecommunications

**Competing financial interests:** The authors declare no competing financial interests.

**Reprints and permission** information is available online at http://npg.nature.com/reprintsandpermissions/

**How to cite this article:** Sando, D. *et al.* Large elasto-optic effect and reversible electrochromism in multiferroic BiFeO$_3$. *Nat. Commun.* 7:10718 doi:10.1038/ncomms10718 (2016).

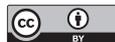